\documentclass[12pt]{article}
\usepackage{amsmath,amssymb,fullpage}
\usepackage{showlabels}
\usepackage[applemac]{inputenc}

\newenvironment{myenumerate}{%

\begin{enumerate}}{\end{enumerate}}

\newcommand{\dproof}{\noindent {Proof.} \quad}
\newcommand{\fproof}{\hfill $\square$ \bigskip}

\newtheorem{definition}{Definition}[section]
\newtheorem{example}{Example}[section]
\newtheorem{theorem}[definition]{Theorem}

\newtheorem{remark}[definition]{ \it Remark}

\newtheorem{proposition}[definition]{Proposition}

\numberwithin{equation}{section}

\def\RB{\mathbb{R}}

\def\FB{\mathbb{F}}

\def\UB{\mathbb{U}}

\def\FC{\mathcal{F}}

\def\EC{\mathcal{E}}
\def\AC{\mathcal{A}}

\def\CC{\mathcal C}

\def\MC{\mathcal M}

\def\1B{\text{1\!\!I}}

\def\tN{\tilde{N}}

\def\tth{\tilde{\theta}}
\def\tvar{\tilde{\varphi}}
\def\tmu{\tilde{\mu}}
\def\hX{\hat{X}}

\def\hu{\hat{u}}

\def\hp{\hat{p}}
\def\hth{\hat{\theta}}
\def\hq{\hat{q}}
\def\hr{\hat{r}}

\def\hpi{\hat{\pi}}

\def\hvar{\hat{\varphi}}
\def\hmu{\hat{\mu}}

\begin{document}

\title{Dynamic robust duality in utility maximization}

\author{
Bernt \O ksendal$^{1,2}$ \and Agn\`es Sulem$^{3,4}$}

\date{24 August 2015}

\footnotetext[1]{
Department of Mathematics, University of Oslo, P.O. Box 1053 Blindern, N--0316 Oslo, Norway,
email: {\tt oksendal@math.uio.no}.
The research leading to these results has received funding from the European Research Council under the European Community's Seventh Framework Programme (FP7/2007-2013) / ERC grant agreement no [228087].}
\footnotetext[2]{Norwegian School of Economics, Helleveien 30, N--5045 Bergen, Norway.}
\footnotetext[3]{INRIA Paris-Rocquencourt, Domaine de Voluceau, Rocquencourt, BP 105, Le Chesnay Cedex, 78153, France, email: {\tt agnes.sulem@inria.fr}}
\footnotetext[4]{Université Paris-Est, F-77455 Marne-la-Vallée, France.}

\maketitle

\begin{abstract}
A celebrated financial application of convex duality theory gives an explicit relation between the following two quantities: 
\begin{myenumerate}
\item
 The optimal terminal wealth $X^*(T) : = X_{\varphi^*}(T)$ of the problem to maximize the expected $U$-utility of the terminal wealth $X_{\varphi}(T)$ generated by admissible portfolios $\varphi(t); 0 \leq t \leq T$ in a market with the risky asset price process modeled as a semimartingale; 
 \item
 The optimal scenario $\frac{dQ^*}{dP}$ of the dual problem to minimize the expected $V$-value of $\frac{dQ}{dP}$ over a family of equivalent local martingale measures $Q$, where  $V$ is the convex conjugate function of the concave function $U$.
 \end{myenumerate}
 
In this paper we consider markets modeled by Itô-L\'evy processes.  In the first part we use the maximum principle in stochastic control theory to extend the above relation to a \emph{dynamic} relation, valid for all $t \in [0,T]$. We prove in particular that the optimal adjoint process for the  primal problem coincides with the optimal density process, and that the optimal adjoint process for the dual problem coincides with the optimal wealth process; $0 \leq t \leq T$. In the terminal time case $t=T$ we recover the classical duality connection above.\\
We get moreover an explicit relation between the optimal portfolio $\varphi^*$ and the optimal measure $Q^*$. 
We also obtain that the existence of an optimal scenario is equivalent to the replicability of a related $T$-claim.\\
In the second part we present robust (model uncertainty) versions of the optimization problems in (i) and (ii), and we prove a similar dynamic relation between them. In particular, we show how to get from the solution of one of the problems to the other.
We illustrate the results with explicit examples.

\end{abstract}



\section{Introduction}\label{sec1}

The purpose of this paper is to use stochastic control theory to obtain new results on the connections between the primal, utility maximization portfolio problem and its convex  dual, both in the non-robust and the robust (worst case  scenario/multiple-priors) setting.
This  approach allows us to get  more detailed information about the connection between the primal and the dual problem. 
In particular, we show that the optimal wealth process 
of the primal problem coincides with the optimal adjoint process 
for the dual problem. This generalizes results that have been obtained earlier by using convex duality theory.


First, let us briefly recall the main results from the duality method in utility maximization, as presented in e.g. \cite{KS}:
Let $U:[0,\infty] \rightarrow \RB$ be a given utility function, assumed to be strictly increasing, strictly   concave,   continuously differentiable $(C^1)$ and satisfying the Inada conditions:
\begin{align*}
U'(0) & = \lim_{x \rightarrow 0^+} U'(x) = \infty \\
U'(\infty) & = \lim_{x \rightarrow \infty} U'(x) = 0.
\end{align*}
Let $S(t) = S(t,\omega) \; ; \; 0 \leq t \leq T$, $\omega \in \Omega$, represent the discounted unit price of a risky asset at time $t$ in a financial market. We assume that $S(t)$ is a semimartingale on a filtered probability space $(\Omega, \FC, \FB:=\{\FC_t\}_{0 \leq t \leq T}, P)$. Let  $\varphi(t)$ be an $\FB$-predictable portfolio process, giving the number of units held of the risky asset at time $t$. If $\varphi(t)$ is  self-financing, the corresponding wealth process $X(t) := X^x_\varphi(t)$ is given by
\begin{equation}\label{eq1.1}
X(t) = x + \int_0^t \varphi(s) dS(s) \; ; \; 0 \leq t \leq T,
\end{equation}
where $T \geq 0$ is a fixed terminal time and $x > 0$ is the initial value of the wealth. We say that $\varphi$ is {\it admissible} and write $\varphi \in \AC$ if the integral in \eqref{eq1.1} converges and
\begin{equation}\label{eq1.2}
X_\varphi^x(t) > 0 \text{ for all } t \in [0,T], \text{ a.s.}.
\end{equation}

The classical optimal portfolio problem  is to find $\varphi^* \in \AC$ (called an optimal portfolio) such that
\begin{equation}\label{eq1.3}
u(x) := \sup_{\varphi \in \AC} E [U(X^x_\varphi(T))] = E[U(X^x_{\varphi^*}(T))].
\end{equation}

The duality approach to this problem is as follows: Let
\begin{equation}\label{eq1.4}
V(y) := \sup_{x > 0} \{U(x) - xy\} \; ; \; y > 0
\end{equation}
be the {\it convex conjugate function} of $U$. Then it is well-known that $V$ is strictly convex, decreasing, $C^1$ and satisfies
\begin{equation}\label{eq1.5}
V'(0) = - \infty, \; V'(\infty) = 0,\;  V(0) = U(\infty) \;  \text{ and } V(\infty) = U(0).
\end{equation}
Moreover,
\begin{equation}\label{eq1.6}
U(x) = \inf_{y > 0} \{ V(y) + xy\} \; ; \; x > 0,
\end{equation}
and
\begin{equation}\label{eq1.7}
U'(x) = y \Leftrightarrow x = -V'(y).
\end{equation}

Let $\MC$ be the set of probability measures $Q$ which are equivalent local martingale measures (ELMM), in the sense that $Q$ is equivalent to $P$ and $S(t)$ is a local martingale with respect to $Q$.
We assume that  $\MC \ne \emptyset $, which means absence of arbitrage opportunities on the financial market.
The dual problem to \eqref{eq1.3} is for given $y >0$  to find $Q^* \in \MC$ (called an optimal scenario measure) such that
\begin{equation}\label{eq1.8}
v(y):= \inf_{Q \in \MC} E \left[ V \left( y \frac{ dQ}{dP}\right) \right] = E \left[ V\left(y \frac{dQ^*}{dP}\right)\right].
\end{equation}

One of the main results in \cite{KS} is that, under some conditions, $\varphi^*$ and $Q^*$ both exist and they are related by
\begin{equation}\label{eq1.9}
U'(X_{\varphi^*}^x(T)) =y  \frac{dQ^*}{dP} \quad \text{ with }  y = u'(x)
\end{equation}
i.e.
\begin{equation}\label{eq1.10}
X_{\varphi^*}^x(T) = - V' \left(y \frac{dQ^*}{dP}\right) \quad \text{ with }  x = -v'(y).
\end{equation}

In this paper we extend this result to a \emph{dynamic} identity between processes by using stochastic control theory. We work in the slightly more special market setting with a risky asset price $S(t)$ described by an Itô-Lévy process. This enables us to use the machinery of the maximum principle and backward stochastic differential equations (BSDE) driven by Brownian motion $B(t)$ and a compensated Poisson random measure $\tN(dt,d\zeta) \; ; \; t \geq 0 \; ; \; \zeta \in \RB_0 := \RB \backslash \{0\}$.  (We refer to e.g. \cite{OS1} for more information about the maximum principle). Our approach has the advantage that it gives a dynamic relation between the optimal scenario in the dual formulation and the optimal portfolio in the primal formulation:

In particular, in Section 3 we prove that 
\begin{equation}
\hat{X}(t) = \hat{p}_2(t) ; t \in [0,T]
\end{equation} 
where $\hat{X}(t)$ is the optimal wealth process and $\hat{p}_2(t)$ is the adjoint process for the dual problem. When $t=T$ this gives the classical duality result above, namely
\begin{equation}
\hat{X}(T) = - V' \left(y \frac{dQ^*}{dP}\right) (= \hat{p}_2(T) ).
\end{equation}
A similar result is obtained for the optimal density process for the dual problem.
 %
As a step on the way, we prove in Section \ref{sec2} a result of independent interest, namely that the existence of an optimal scenario is equivalent to the replicability of a related $T$-claim.

Then in Section 4 we extend the discussion to \emph{robust (model uncertainty)} optimal portfolio problems.   More precisely, we formulate robust versions of the primal problem \eqref{eq1.3} and of the dual problem  \eqref{eq1.8}, we establish a corresponding dynamic identity between processes  and we show explicitly how to get from the solution of one to the solution of the other.

This  paper addresses duality of robust utility maximization problems entirely by means of stochastic control methods, but  there are  several papers of related interest based on convex duality methods, see e.g. the survey paper \cite{FSW} and the references therein. We also refer the reader to  \cite{Q}  where  the author uses convex duality to study utility maximization under model uncertainty (multiple prior) and obtains a BSDE characterization of the optimal wealth process in markets driven by Brownian motion. 
 In  \cite{G},  a robust dual characterization of the robust primal utility maximization problem is obtained by convex duality methods. The dual formulation obtained is similar to ours, but there is no BSDE connection.\\
 None of the above papers deal with a \emph{dynamic} duality.


\section{Dynamic duality in utility maximization}
\subsection{Optimal portfolio, optimal scenario and replicability} \label{sec2}

We now specialize the setting described in Section \ref{sec1} as follows:
Suppose the financial market has a risk free asset with unit price $S_0(t)=1$ for all $t$ and a risky asset with price $S(t)$ given by
\begin{equation}\label{eq2.1}
\begin{cases}
dS(t)  = \displaystyle S(t^-)\left(b(t)dt+ \sigma(t) dB(t) + \int_\RB \gamma(t,\zeta) \tN (dt,d\zeta) \right)\; ; \; 0 \leq t \leq T \\
S(0)  >0
\end{cases}
\end{equation}
where $b(t), \sigma(t)$ and $\gamma(t,\zeta)$ are predictable processes satisfying $\gamma > -1$ and
\begin{equation}\label{eq2.2}
E \left[ \int_0^T \left\{ | b(t)| + \sigma^2(t) + \int_\RB \gamma^2(t,\zeta) \nu(d\zeta) \right\} dt \right] < \infty.
\end{equation}
Here $B(t)$ and $\tN(dt,d\zeta):= N(dt,d\zeta) -\nu(d\zeta)dt$ is a Brownian motion and an independent  compensated Poisson random measure, respectively, on a filtered probability space $(\Omega,   \FC, \FB:=\{\FC_t\}_{0 \leq t \leq T}, P)$ satisfying the usual conditions,  $P$ is a reference probability measure and $\nu$ is the Lévy measure of $N$.

Let $\varphi(t)$ be a self financing portfolio and let  $X(t) := X^x_ \varphi(t) $ be the corresponding wealth process  given by
\begin{equation}\label{eq2.3}
\begin{cases}
dX(t) = \varphi(t) S(t^-)\left[ b(t) dt + \sigma(t) dB(t) + \int_\RB \gamma(t,\zeta) \tN(dt,d\zeta)\right] \; ; \; 0 \leq t \leq T \\
X(0) = x > 0.
\end{cases}
\end{equation}

\begin{definition}[Admissible Portfolios]
Let $\varphi$  be an $\FB$-predictable, $S$-integrable  process. 
We say that $\varphi$ is admissible if  
\begin{align}
&X
(t) > 0 \text{ for all } t \in [0,T], \text{ a.s.}  \nonumber \\
&E \left[ \int_0^T \varphi(t)^2 S(t)^2 \left\{   b(t) ^2 +\sigma^2(t) +  \int_\RB \gamma^2(t,\zeta) \nu (d \zeta) \right\} dt \right] < \infty, \nonumber \\
&  E [ \int_0^T  | X(t) | ^{2}dt] < \infty \label{eq2.A1} \\
&E [  U'(X(T)) ^{2}]   < \infty. \label{eq2.A2}
\end{align}
\end{definition}
We denote by  $\AC$ the set of admissible portfolios. 
Conditions  \eqref{eq2.A1}, \eqref{eq2.A2} are needed for the application of the maximum principles. See Appendix A.

As in \eqref{eq1.3}, for given $x>0$,  we want to find $\varphi^* \in \AC$ such that
\begin{equation}\label{eq2.4}
u(x) := \sup_{\varphi \in \AC} E [U(X_\varphi^x(T))] = E[U(X_{\varphi^*}^x(T))].
\end{equation}
We consider the family $\MC$ of equivalent local martingale measures (ELMM) that can be represented by means of  the family of positive measures $Q = Q_\theta$ of the form
\begin{equation}\label{eq2.5}
dQ_\theta(\omega) = G_\theta(T) dP(\omega) \text{ on } \FC_T,
\end{equation}
where
\begin{equation}\label{eq2.6}
\begin{cases}
dG_\theta (t) = G_\theta(t^-) \left[ \theta_0(t) dB(t) + \int_\RB \theta_1(t,\zeta) \tN (dt,d\zeta)\right] \; ; \; 0 \leq t \leq T \\
G_\theta(0) = y > 0,
\end{cases}
\end{equation}
and $\theta = (\theta_0, \theta_1)$ is a predictable process satisfying the conditions
\begin{equation}\label{eq2.7}
E\left[ \int_0^T \left\{ \theta^2_0(t) + \int_\RB \theta^2_1 (t,\zeta) \nu (d\zeta) \right\} dt \right] < \infty, \; \theta_1(t,\zeta) > -1 \;\; \text{ a.s.} 
\end{equation}
and
\begin{equation}\label{eq2.8}
b(t) + \sigma(t) \theta_0(t) + \int_\RB \gamma(t,\zeta) \theta_1 (t,\zeta) \nu (d\zeta) = 0 \; ; \; t \in [0,T].
\end{equation}
If $y=1$ this condition implies  that $Q_\theta$ is an ELMM for this market. See e.g. \cite[Chapter 1]{OS1}. 
\begin{remark}
The set  $\MC$ with $y=1$ is contained in the set ELMM. Note, however, that there are ELMM's which are not of the above form. But $\MC$ is the family we choose to work with, and all our results are proved for this family of measures. \end{remark}
We let $\Theta$ denote the set of all $\FB$-predictable processes $\theta = (\theta_0, \theta_1)$ satisfying \eqref{eq2.7}-\eqref{eq2.8}. \\
The dual problem corresponding to \eqref{eq1.8} is for given $y >0$  to find $\hth \in \Theta$ and $v(y)$ such that
\begin{equation}\label{eq2.9}
- v(y) := \sup_{\theta \in \Theta} E [-V(G^y_\theta(T))] = E[-V(G^y_{\hth}(T))].
\end{equation}

We will use two  stochastic maximum principles for stochastic control to study the problem \eqref{eq2.9} and relate it to \eqref{eq2.4}. We refer to Appendix A for a presentation of these principles and to \cite{QS} for more information about backward stochastic differential equations (BSDEs) with jumps.

We recall the existence and uniqueness result for BSDEs with jumps,  due to Tang and Li (1994) (see \cite{Tang}).
If  $T>0$, $F \in L^2({\cal F}_T)$, and  $g$ is a Lipschitz driver, then  there exists a unique solution $\in S^2 \times H^2 \times H_{\nu}$ of the BSDE with jumps
\begin{align}\label{BSDE}
dp(t) &= -g(t,p(t),q(t),r(t,\cdot))dt + q(t)dB(t) + \int_{\mathbb{R}} r(t,\zeta) \tN(dt,d\zeta)\; ; \; 0 \leq t \leq T \nonumber\\
p(T) &= F,
\end{align}
 where 
\begin{itemize}
\item  $S^2$ is the set of real-valued c\`adl\`ag adapted 
 processes $\phi$ with $E(\sup_{0\leq t \leq T} |\phi_t |^2) <  \infty.$
 \item $H^2$ is the set of 
real-valued predictable processes $\phi$ such that $ E \left[(\int_0 ^T \phi_t ^2 dt)\right] < \infty,$ 
\item $H_{\nu}^2$ is  the set of  predictable processes $\ell$  such that $E\left[( \int_0 ^T 
(\int_{\mathbb{R}}  |\ell(t,\zeta) |^2 \nu(d\zeta) )
 \,dt )  \right]< \infty.$
 \end{itemize}
From now on, when we say that a process triple $(p(t),q(t),r(t,\zeta))$ satisfies a BSDE of the form \eqref{BSDE}, 
it is tacitly understood that $(p,q,r) \in S^2 \times H^2 \times H_{\nu}$.

We first prove two auxiliary results, the first of which may be regarded as a special case of Proposition 4.4 in \cite{OS5}.
\begin{proposition}[Primal problem and associated constrained FBSDE]\label{prop2.1}
Let $\hat{\varphi}$  in $\AC$. Then $\hat{\varphi} $ is optimal for the primal problem \eqref{eq2.4} if and only if the 
(unique) solution  $X$, $(\hp_1,\hq_1,\hr_1)$ in $ S^2 \times H^2 \times H_{\nu}^2$  of the FBSDE consisting of the  SDE \eqref{eq2.3} and the BSDE
\begin{equation}\label{equa2.13}
\begin{cases}
\displaystyle d\hp_1(t) =  \hq_1(t) dB(t) + \int_\RB \hr_1(t,\zeta) \tN(dt,d\zeta) \; ; \; 0 \leq t \leq T \\
 \hp_1(T) =  U'(X_{\hat{\varphi}}^x(T))
 \end{cases}
 \end{equation}
satisfies the equation
\begin{equation}\label{eq2.13b}
b(t)\hp_1(t) + \sigma(t) \hq_1(t) + \int_\RB \gamma(t,\zeta) \hr_1 (t,\zeta) \nu (d\zeta) = 0 \; ; \; t \in [0,T].
\end{equation}

\end{proposition}
\dproof
(i) \ 
The Hamiltonian corresponding to the primal problem is given by 
\begin{equation}
H_1(t,x, \varphi, p,q,r) = \varphi S(t^-) ( b(t) p + \sigma(t) q + \int_\RB \gamma(t, \zeta)  r(\zeta) \nu(d\zeta) ).
\end{equation}
Assume $\hat{\varphi} \in \AC$ is optimal for the primal problem \eqref{eq2.4}.
Then by the necessary maximum principle (Theorem \ref{thmA.2}), we have 
$$
\frac{\partial H_1}{ \partial \varphi}(t,x,\varphi, \hp_1(t), \hq_1(t), \hr_1(t, \cdot)) \mid_{\varphi = \hat{\varphi}(t)} = 0, 
$$
where $(\hp_1, \hq_1, \hr_1)$ satisfies \eqref{equa2.13}, since 
$
\frac{\partial H_1}{ \partial x}(t,x,\varphi, \hp_1(t), \hq_1(t), \hr_1(t, \cdot)) = 0.$
This implies \eqref{eq2.13b}.\\

(ii) \  Conversely, suppose the solution  $(\hp_1, \hq_1, \hr_1)$ of the BSDE \eqref{equa2.13} satisfies \eqref{eq2.13b}.
Then $\hat{\varphi}$, with the associated $(\hp_1, \hq_1, \hr_1)$ satisfies the conditions for the sufficient maximum principle 
(Theorem \ref{thA1}) with the additional feature of a constraint. See \eqref{eq2.20a} below. We conclude that $\hat{\varphi}$ is optimal. 
\fproof

  \begin{remark}\label{rem23}
  The BSDE \eqref{equa2.13} is linear, and hence it is well known that it has a unique solution $(p,q,r)$ for every choice of $X_{\varphi}^x(T)$. See e.g. \cite{QS},\cite{R}. We are seeking $ \hat{\varphi}$ such that the corresponding solution $(\hp,\hat{q},\hat{r})$ of \eqref{equa2.13} also satisfies \eqref{eq2.13b}.
  \end{remark}
 
 \begin{remark}\label{rem24}
By \eqref{equa2.13} we have 
$\hp_1(t) = E [U'(X_{\hat{\varphi}}^x(T)) \mid \FC_t]>0$ for all $t$ in $[0,T]$,
and if we divide equation \eqref{eq2.13b} throughout by $\hp_1(t)$ we get
\begin{equation} \label{eq2.13c}
b(t) + \sigma(t) \hth_0(t) + \int_{\RB} \gamma(t,\zeta) \hth_1(t,\zeta) \nu(d\zeta) = 0 \; ; \; t \in [0,T],
\end{equation}
where
\begin{equation} \label{eq2.13d}
\hth_0(t):  = \frac{\hq_1(t)}{\hp_1(t)}  \; ;  \;\; 
 \hth_1(t,\zeta) : = \frac{\hr_1(t,\zeta)}{\hp_1(t)} \; , \; t \in [0,T].
\end{equation}
By the Girsanov theorem this is saying that if we define the measure
$ Q_{(\hth_0,\hth_1)} $ as in \eqref{eq2.5},\eqref{eq2.6} with $y=1$, then $ Q_{(\hth_0,\hth_1)} $ is an ELMM for the market described by \eqref{eq2.1}.
 \end{remark}
  We now turn to the dual problem \eqref{eq2.9}:

\begin{proposition}[Dual problem and associated constrained FBSDE]    \label{prop2.2}
Let  $\hth \in \Theta$.  Then  $\hth$ is an optimal scenario for the dual problem \eqref{eq2.9} if and only if 
the solution $G_{\hth}$, $(\hp_2, \hq_2, \hr_2) $  in  $ S^2 \times H^2 \times H_{\nu}^2$ of the FBSDE consisting of the FSDE \eqref{eq2.6} and BSDE
  \begin{equation}\label{eq2.32}
   \begin{cases}
   d \hp_2(t) & = \displaystyle K(\hq_2, \hr_2)(t) \left[ b(t) dt + \sigma(t) dB(t) + \int_\RB \gamma(t,\zeta) \tN(dt,d\zeta)\right] \\
   \hp_2 (T)& = - V'(G^y_\theta(T))
   \end{cases}
   \end{equation}
   where
   \begin{equation}\label{eq2.19}
   K(q,r)(t) := \frac{q(t)}{\sigma(t)} \chi_{\sigma(t) \neq 0} + \frac{r(t,\zeta)}{\gamma(t,\zeta)} \chi_{\sigma(t) = 0, \gamma(t,\zeta) \neq 0}
   \end{equation}
 
also satisfies 
  \begin{equation}\label{eq2.17}
  - \hq_2(t) \gamma(t,\zeta) + \sigma(t) \hr_2(t,\zeta) = 0 \; ; \; 0 \leq t \leq T.
  \end{equation}

\end{proposition}
\dproof
%
%
%
 We may regard the problem \eqref{eq2.9} as a stochastic control problem in the control process $\theta$ with the constraint \eqref{eq2.8}. To solve this problem we use the well-known Lagrange multiplier technique. Thus we define the Hamiltonian $H_2^L$ by 
  \begin{align}\label{eq2.20a}
  H_2^L(\theta_0, \theta_1,L) := g \theta_0 q + g \int_\RB \theta_1(\zeta) r(\zeta) \nu(d\zeta)
    + L(t) \left( b(t) + \sigma(t) \theta_0 + \int_\RB \gamma(t,\zeta) \theta_1(\zeta) \nu (d\zeta)\right),
   \end{align}
   where $L(t)$ is the Lagrange multiplier process.
      Maximizing $H_2^L$ over all $\theta_0$ and $\theta_1$ gives the following first order conditions
  $$
   gq + L(t) \sigma(t) = 0; \; \;
   gr(\cdot) + L(t) \gamma(t,\cdot) = 0.
  $$ 
   Since $g = G_\theta(t) \neq 0$, we can write these as follows:
   \begin{equation}\label{eq2.31}
   q(t) = - \frac{L(t)}{G_\theta(t)} \sigma(t); \;\; 
   r(t,\zeta) = - \frac{L(t)}{G_\theta(t)} \gamma(t,\zeta).
   \end{equation}
   The adjoint equation becomes:
      \begin{equation}\label{eq2.33}
   \begin{cases}
   dp(t) &= \displaystyle - \frac{L(t)}{G_\theta(t)} \left[ \left\{ - \theta_0(t) \sigma(t) - \int_\RB \theta_1(t,\zeta) \gamma(t,\zeta) \nu (d\zeta) \right\} \right.dt \\
   &\left. + \sigma(t) dB(t) + \int_\RB \gamma(t,\zeta) \tN (dt, d\zeta)\right]  \; ; \; 0 \leq t \leq T \\
   p(T) & = - V' (G_\theta(T)).
   \end{cases}
   \end{equation}
   In view of \eqref{eq2.8} this can be written
   \begin{equation}\label{eq2.34}
   \begin{cases}
   \displaystyle
   dp(t) = - \frac{L(t)}{G_\theta(t)} \left[ b(t) dt + \sigma(t) dB(t) + \int_\RB \gamma(t,\zeta) \tN(dt,d\zeta)\right] \; ; \; 0 \leq t \leq T \\
   \displaystyle p(T) = - V'(G_\theta(T))
   \end{cases}
   \end{equation}
   Note that
   \begin{equation}\label{eq2.35}
   \text{If } \sigma(t) \neq 0 \text{ then } - \frac{L(t)}{G_\theta(t)} = \frac{q(t)}{\sigma(t)}
   \end{equation}
   \begin{equation}\label{eq2.36}
   \text{If } \gamma(t,\zeta) \neq 0 \text{ then } - \frac{L(t)}{G_\theta(t)} = \frac{r(t,\zeta)}{\gamma(t,\zeta)}
   \end{equation}
If $\sigma(t) = \gamma(t,\zeta) = 0$, then by \eqref{eq2.31}  we have $q(t) = r(t,\zeta) = 0$ and hence 
we have  $dp(t) = 0$. Therefore, with $K(q,r)(t)$ defined as in \eqref{eq2.19}, we get 
%
by \eqref{eq2.34}
\begin{equation}\label{eq2.38}
\begin{cases}
\displaystyle dp(t) = K(q,r)(t)\left[ b(t)dt + \sigma(t) dB(t) + \int_\RB \gamma(t,\zeta) \tN (dt,d\zeta)\right] \; ; \; 0 \leq t \leq T \\
p(T) = - V'(G_\theta(T)).
\end{cases}
\end{equation}
By combining the two equations of \eqref{eq2.31} we get 
\eqref{eq2.17}. 
 This completes the proof of the necessary part. 
   \fproof

  The sufficient part follows from the fact that the functions $g \rightarrow - V(g)$ and
 $$g \rightarrow \sup_{\theta_0,\theta_1} {H}_2^L(t,g,\theta_0,\theta_1,\hp_2(t), \hq_2(t), \hr_2(t,\cdot)) $$
 are concave.
 \fproof

  We deduce as a by-product the following results of independent interest which relates the existence of a solution of the dual problem to the replication of a related $T$-claim.

  \begin{proposition}\label{th2.1}
  For given $y>0$ and  $\hth \in \Theta$ the following are equivalent:
  \begin{myenumerate}
  \item 
  $$\sup_{\theta \in \Theta} E [-V(G_\theta^y(T))] = E[-V(G_{\hth}^y (T))] < \infty.$$
  \item 
  The claim $F:= - V' (G_{\hth} ^y(T))$ is replicable, with initial value $x = \hp_2(0)$, where $(\hp_2, \hq_2, \hr_2)$ solves 
    \begin{equation}\label{e2.18}
  \begin{cases}
 \displaystyle  d\hp_2(t) = K(\hq_2,\hr_2)(t) \left[ b(t) dt + \sigma(t) dB(t) + \int_\RB \gamma(t,\zeta) \tN (dt,d\zeta)\right] \; ; \; 0 \leq t \leq T \\
  \hp_2(T) = -V' (G_{\hth}^y(T)).
  \end{cases}
  \end{equation}
  \end{myenumerate}
  Moreover, if (i) or (ii) holds, then
  \begin{equation}\label{eq221}
 \hat{\varphi}(t) : = \frac{K(\hq_2,\hr_2)(t)}{S(t^-)}
  \end{equation}
  is a replicating portfolio for $F:= -V'(G_{\hth}^y(T))$, where $(\hp_2,\hq_2,\hr_2)$ is the solution of the BSDE \eqref{e2.18}.
 \end{proposition}
  \dproof
  (i) $\Rightarrow$ (ii): We have already proved that (i) implies \eqref{e2.18}.
This equation  states that the contingent claim $F:= - V'(G_{\hth}^y(T))$ is replicable, with replicating portfolio $\hat{\varphi}(t)$ given by \eqref{eq221} 
  and initial value $x = \hp_2(0)$. Note that $\hp_2(t) > 0$ for all  $t$, since $V$ is strictly decreasing so  $- V'(G_{\hth}^y(T)) > 0$. 

   (ii) $\Rightarrow$ (i): Suppose 
    $F:= - V' (G_{\hth}^y(T))$ is replicable with initial value $x = \hp_2(0)$, and let $\varphi \in \AC$ be a replicating portfolio. Then $X(t) = X^x_{\varphi}(t)$ satisfies the equation
  \begin{equation}\label{eq2.20}
  \begin{cases}
 \displaystyle  dX(t) = \varphi(t) S(t^-) \left[ b(t) dt + \sigma(t) dB(t) + \int_\RB \gamma(t,\zeta) \tN (dt,d\zeta)\right] \; ; \; 0 \leq t \leq T \\
  X(T) = - V' (G_\theta^y(T)).
  \end{cases}
  \end{equation}
  Define
  \begin{equation}\label{eq2.21}
  \hp(t) := X(t), \hq(t) := \varphi(t) \sigma(t)  S(t^-) \text{ and } \hr(t,\zeta) := \varphi(t)  \gamma(t,\zeta) S(t^-).
  \end{equation}
  They satisfy the relation \eqref{eq2.17}. 
  Moreover, by \eqref{eq2.21} we get
  \begin{equation}\label{eq2.21a}
  \varphi(t)S(t^-) = K(q,r)(t).
  \end{equation}
  Therefore, from \eqref{eq2.20} we get  that 
  $(\hp,\hq, \hr)$ satisfies the BSDE
  \begin{equation}\label{eq2.22}
  \begin{cases}
  \displaystyle d\hp(t) = K(\hq,\hr)(t) \left[ b(t)dt + \hq(t) dB(t) + \int_\RB \hr(t,\zeta) \tN (dt,d\zeta)\right] \; ; \; 0 \leq t \leq T \\
  \hp(T) = - V' (G_{\hth}^y(T)).
  \end{cases}
  \end{equation}
 We conclude that $ \hp(t) = X(t) = \hp_2(t)$.  
  Hence (i) holds, by Proposition \ref{prop2.2}.
 
 The last statement follows from \eqref{eq2.21a}.
  \fproof

\subsection{Relations between optimal scenario and optimal portfolio}\label{sec3}

We proceed to show that the method above actually gives a connection between an optimal scenario  $\hth \in \Theta$ for the dual  problem \eqref{eq2.9} and an optimal portfolio $\hvar \in \AC$ for the primal problem \eqref{eq2.4}.

\begin{theorem}\label{th3.1}

\noindent {\bf a)} Suppose $\hvar \in \AC$ is optimal for the primal  problem \eqref{eq2.4}.  \\
Let $(\hp_1(t), \hq_1(t), \hr_1(t,\zeta))$ be the associated adjoint processes, solution of the constrained BSDE \eqref{equa2.13}-\eqref{eq2.13b}.
Define
\begin{equation}\label{eq3.2}
\hth_0(t) = \frac{\hq_1(t)}{\hp_1(t^-)}, \;\;\; \hth_1(t,\zeta) = \frac{\hr_1(t,\zeta)}{\hp_1(t^-)}.
\end{equation}
Suppose 
\begin{equation}\label{eq3.2b}
E[ \int_0^T \{\hth_0^2(t) +  \int_\RB \hth_1^2(t,\zeta)\nu(d\zeta) \}dt] < \infty; \; \; \hth_1 >-1.
\end{equation}
Then $\hth = (\hth_0, \hth_1) \in \Theta$ is optimal for the dual  problem \eqref{eq2.9} with initial value $y = \hp_1(0)$. 
Moreover, with $y = \hp_1(0)$, 
\begin{equation}\label{34n}  G_{\hth}^y(t) =\hp_1(t) ; \quad t \in [0,T] .
\end{equation}
  In particular 
\begin{equation}\label{eq3.3}
G_{\hth}^y(T) = U'(X_{\hvar}^x(T)).
\end{equation}

\noindent {\bf b)} Conversely, suppose $\hth = (\hth_0, \hth_1) \in \Theta$ is optimal for the dual  problem \eqref{eq2.9}.
 Let $(\hp_2(t), \hq_2(t), \hr_2(t,\zeta))$ be the associated adjoint processes,  solution of the  BSDE
\eqref{eq2.32} with the constraint \eqref{eq2.17}.
Suppose the portfolio
 \begin{equation}\label{eq3.5}
 \hat{\varphi}(t) : = \frac{K(\hq_2,\hr_2)(t)}{S(t^-)}
  \end{equation}
is admissible. Then $\hvar$ 
is an optimal portfolio for the primal problem \eqref{eq2.4}  with initial value $x = \hp_2(0)$.
Moreover, with $x = \hp_2(0)$, 
\begin{equation}\label{38n}
X_{\hvar}^x(t) =  \hp_2(t) ;  \quad t \in [0,T]. 
\end{equation}
In particular 
\begin{equation}\label{eq3.6}
X_{\hvar}^x(T) = - V' (G_{\hth}^y(T)).
\end{equation}
\end{theorem}
\fproof

\dproof
 {\bf a)} Suppose  $\hvar$ is optimal for problem \eqref{eq2.4} with initial value $x$. Then, by Proposition 
 \ref{prop2.1}, the adjoint processes $\hp_1(t), \hq_1(t), \hr_1(t,\zeta)$ for Problem \eqref{eq2.4} satisfy \eqref{equa2.13}-\eqref{eq2.13b}.
Consider the process $\hth(t)$ defined in \eqref{eq3.2} and suppose 
 \eqref{eq3.2b} holds. 
Then $\hth \in \Theta$ and \eqref{equa2.13} can be written
\begin{equation}\label{eq3.10}
\begin{cases}
\displaystyle d\hp_1(t) = \hp_1(t^-) \left[ \hth_0(t) dB(t) + \int_\RB \hth_1 (t,\zeta) \tN (dt, d\zeta)\right] \\
\hp_1(T) = U'(X_{\hvar}^x(T)).
\end{cases}
\end{equation}
Therefore $\hp_1(t) \equiv G_{\hth}^y(t) $ (see \eqref{eq2.6}) if we put  $y :=\hp_1(0)>0$, 
 and we have, by \eqref{eq1.7}
\begin{equation}\label{eq3.11}
U'(X_{\hvar}^x(T)) = G_{\tth}^y(T), \text{ i.e. } X_{\hvar}^x(T) = - V'(G_{\tth}^y(T)).
\end{equation}
Now define
  \begin{equation}\label{eq2.21b}
  \hp_2(t) := X_{\hvar}^x(t), \hq_2(t) := \hvar(t) \sigma(t)  S(t^-) \text{ and } \hr_2(t,\zeta) := \hvar(t)  \gamma(t,\zeta) S(t^-).
  \end{equation}
Then $ (\hp_2, \hq_2,  \hr_2)$ satisfy the conditions of Proposition \ref{prop2.2}      which imply that 
$\hth$ is optimal for problem \eqref{eq2.9}.


\medskip

\noindent {\bf b)} Suppose $\hth  \in \Theta$  is optimal for problem \eqref{eq2.9} with initial value $y$. Let 
$\hp_2(t), \hq_2(t), \hr_2(t,\cdot)$ be the associated adjoint processes, solution of the BSDE \eqref{eq2.32} with
the constraint  \eqref{eq2.17}. 
Then
they  satisfy the equation
\begin{equation}\label{eq3.12}
\begin{cases}
\displaystyle d\hp_2(t) = K(\hq_2,\hr_2)(t) \left[ b(t) dt + \sigma(t) dB(t) + \int_\RB \gamma(t,\zeta) \tN(dt,d\zeta)\right] \\
\hp_2(T) = - V'(G_{\hth}(T)).
\end{cases}
\end{equation}
Define
\begin{equation}\label{eq3.13}
\tvar(t)  :=  \hat{\varphi}(t) : = \frac{K(\hq_2,\hr_2)(t)}{S(t^-)},
\end{equation}
and assume $\tvar(t) $ is admissible. 
Then
$ \hp_2(t) \equiv X_{\tvar}^x(t) $ for $x = \hp_2(0)$.
In particular
\begin{equation}\label{eq3.14}
X_{\tvar}^x(T) = - V'(G_{\hth}^y(T)),  \text{ i.e. } G_{\hth}^y(T) = U' (X_{\tvar}^x(T)).
\end{equation}
Therefore $G_{\hth}^y(t)= G_{\hth}(t)$ satisfies the equation
\begin{equation}\label{eq3.15}
\begin{cases}
\displaystyle dG_{\hth}(t) = G_{\hth}(t^-) \left[ \hth_0(t) dB(t)  +  \int_\RB \hth_1(t,\zeta) \tN(dt,d\zeta)\right] \; ; \; 0 \leq t \leq T \\
G_{\hth}(T) = U'(X_{\tvar}^x(T)).
\end{cases}
\end{equation}
Define now
\begin{equation}\label{eq3.16} p_1(t) := G_{\hth}(t), q_1(t) := G_{\hth}(t) \hth_0(t), r_1(t,\zeta) := G_{\hth}(t) \hth_1(t,\zeta).
\end{equation}
Then by \eqref{eq3.15} $(p_1,q_1,r_1)$ solves the BSDE
\begin{equation}\label{eq3.17}
\begin{cases}
\displaystyle dp_1(t) = q_1(t) dB(t) + \int_\RB r_1(t,\zeta) \tN(dt,d\zeta) \; ; \; 0 \leq t \leq T \\
p_1(T) = U'(X_{\tvar}^x(T)).
\end{cases}
\end{equation}
Moreover, since $\hth \in \Theta$, it satisfies \eqref{eq2.8}, that is
\begin{equation}
\label{eq3.18}
b(t) + \sigma(t) \hth_0(t) + \int_\RB \gamma(t,\zeta) \hth_1(t,\zeta) \nu (d\zeta) = 0 \; ; \; 0 \leq t \leq T \end{equation}
i.e., $(p_1,q_1,r_1)$ satisfies the equation
\begin{equation}\label{eq3.19}
b(t) + \sigma(t) \frac{q_1(t)}{p_1(t)} + \int_\RB \gamma(t,\zeta) \frac{r_1(t,\zeta)}{p_1(t)} \nu (d\zeta) = 0 \; ; \; 0 \leq t \leq T.
\end{equation}
It follows from Proposition \ref{prop2.1} that $\hvar := \tvar$ is an optimal portfolio for problem \eqref{eq2.4}
with initial value $x = \hp_2(0)$. 
\fproof

\begin{remark}
Conditions of the above theorem  have to be verified in each specific case. 
They hold for examples in 
in Examples  \ref{examples} and \ref{ex2}. 
Note that the integrability condition  in \eqref{eq3.2b} hold whenever the utility function $U$ satisfies the condition
\begin{equation}\label{eq3.18a}
U' \text{  is bounded and bounded away from } 0.
\end{equation}
Indeed this implies that $p_1(t)$ which is equal to  $E [U'(X_{\hat{\varphi}}^x(T)) \mid \FC_t]$ is bounded away from 0 and that ($q_1$, $r_1$) belongs  to $ H^2 \times H_{\nu}$. Therefore $\frac{1}{p_1}$ is bounded and ($\frac{q_1}{p_1}$ , $\frac{r_1}{p_1}$) belong to  $H^2 \times H_{\nu}$. 
Condition \eqref{eq3.18a} does not hold a priori for the most commonly studied utility functions, e.g. the logarithmic or the power functions, but any given utility function can be perturbed slightly such that it holds, simply by modifying it arbitrary near 0 or arbitrary near infinity, if necessary.
\end{remark}
\begin{example}\label{examples}
\rm
As an illustration of Theorem \ref{th3.1} let us apply it to the situation when $\sigma=0$, $\gamma(t,\zeta) = \gamma(t,1) > 0$ and $N(t)$ is the Poisson process with intensity $\lambda > 0$. Then $\nu(d \zeta) = \lambda \delta_1(d\zeta)$, where $\delta_1$ is Dirac measure at 1, and hence
\begin{equation}\label{eq3.20}
\int_\RB \gamma(t, \zeta) \tN (dt, d\zeta) = \gamma(t,1) (dN(t)- \lambda dt) := \gamma (t,1) d \tN(t),
\end{equation}
and \eqref{eq2.1} and  \eqref{eq2.3} become, respectively,
\begin{equation}\label{eq3.20a}
dS(t) = S(t^-)[b(t) dt + \gamma (t,1) d \tN(t)] \; ; \; S(0) > 0
\end{equation}
and
\begin{equation}\label{eq3.20b}
dX(t) = \varphi(t) S(t^-) [b(t) dt + \gamma(t,1) d \tN(t)] \; ; \; X(0) = x > 0.
\end{equation}
Assume that $b(t)$ and $\gamma(t,1)$ are bounded predictable processes and that there exists a constant $C <1$ such that 
\begin{equation}\label{eq3.21a}
\frac{|b(t)|}{\lambda |\gamma(t,1)|} \leq C \; ; \; 0 \leq t \leq T.
\end{equation} 
Then $\Theta$ has just one element $\theta_1(t,1)$, given by
\begin{equation}\label{eq3.21}
\theta_1(t,1) = - \frac{b(t)}{\lambda \gamma (t,1)}
\end{equation}
and hence by \eqref{eq2.8} and the It\^o formula, 
\begin{align}\label{eq3.22}
G^y_{\theta_1}(t) & = y \exp \big( \int_0^t \ln (1 - \frac{b(s)}{\lambda \gamma(s,1)}) d \tN(s) \nonumber \\
 & + \lambda \int_0^t \{ \ln ( 1 - \frac{b(s)}{\lambda \gamma(s,1)}) + \frac{b(s)}{\lambda \gamma (s,1)} \} ds \big) \; ; \; 0 \leq t \leq T.
 \end{align}
 By Theorem \ref{th3.1}b) we get that
 \begin{equation}\label{eq3.23}
 \hvar(t) := \frac{ \hr_2(t,1)}{\gamma(t,1) S(t^-)}\end{equation}
 is an optimal portfolio for the primal problem \eqref{eq2.6}, where $(\hp_2(t), \hr_2(t,1))$ solves the BSDE \eqref{eq2.32}, which in our case gets the form
 \begin{equation}\label{eq3.24}
 \begin{cases}
 d\hp_2(t) & = \frac{\hr_2(t,1)}{\gamma(t,1)} b(t) dt + \hr_2(t,1) d \tN(t) \; ; \; 0 \leq t \leq T \\
 \hp_2 (T) & = - V'(G^y_{\theta_1}(T)).
 \end{cases}
 \end{equation}
  To solve this BSDE we try a solution of the form
 \begin{equation}\label{eq3.25} \hr_2(t,1) = \hp_2(t) \psi(t),
 \end{equation}
 for some predictable process $\psi$, and get the solution
 \begin{align}\label{eq3.26}
 \hp_2(t) & = \hp_2(0) \exp \big( \int_0^t \ln (1 + \psi(s)) d \tN(s)  \nonumber \\
 & + \int_0^t \{ \lambda (\ln (1 + \psi(s)) - \psi(s)) + \frac{b(s)}{\gamma(s,1)} \psi(s) \} ds\big).
 \end{align}
  In particular, if $U(x) = \ln x$, then $V(y) = - \ln y-1$ and $V'(y) = - \frac{1}{y}$. Hence \eqref{eq3.24} implies that
 \begin{align}\label{eq3.27}
 \hp_2(0) & \exp \left( \int_0^T \ln (1 + \psi(s)) d \tN(s) \right.\nonumber \\
 & + \int_0^T \left\{ \lambda ( \ln (1 + \psi(s)) - \psi(s)) + \frac{b(s)}{\gamma(s,1)} \psi(s) \right\}dt \nonumber \\
 & = \frac{1}{y} \exp \left( - \int_0^T \ln \left( 1 - \frac{b(s)}{\lambda \gamma(s,1)}\right) d \tN(s) \right. \nonumber\\
 & \left.  - \lambda \int_0^T \left\{ \ln \left( 1 - \frac{b(s)}{\lambda \gamma(s,1)}\right) + \frac{b(s)}{\lambda \gamma(s,1)}\right\} ds \right).
 \end{align}
 Choose
 \begin{equation}\label{eq3.28}
 \hp_2(0) = \frac{1}{y}
 \end{equation}
 and choose $\psi(s)$ such that the $d\tilde{N}$-integrals of \eqref{eq3.27} coincide, i.e. 
 $$\ln (1 + \psi(s)) = - \ln \left( 1 - \frac{b(s)}{\lambda \gamma(s,1)}\right)$$
 i.e.
 \begin{equation}\label{eq3.29}
 \psi(s) = \frac{b(s)}{\lambda \gamma(s,1) - b(s)}.
 \end{equation}
Then we see that also the $ds$-integrals coincide, i.e. 
 $$\lambda(\ln (1 + \psi(s)) - \psi(s)) + \frac{b(s)}{\gamma(s,1)} \psi(s) = - \lambda \left( \ln \left( 1 - \frac{b(s)}{\lambda \gamma(s,1)}\right) + \frac{b(s)}{\lambda \gamma(s,1)} \right).$$
 Hence, with this choice of $\psi$, we see that
 $d(\hat{p}_2(t)) = d ( \frac{1}{G_{\theta_1}^y})(t).$
 The process 
 $$\hp_2 = \frac{1}{G_{\theta_1}^y}\; ; \; \hr_2 = \hp_2 \psi $$
 with $\psi$ given by \eqref{eq3.29} 
 solves BSDE \eqref{eq3.24}. Moreover \eqref{eq2.17} holds trivially.
 We conclude by \eqref{eq3.23} that the optimal portfolio $\hvar(t)$ for problem \eqref{eq2.6} with $U(x) = \ln x$ is 
 \begin{equation}\label{eq3.31}
 \hvar(t) = \frac{\hp_2(t) b(t)}{\gamma(t,1) S(t^-)(\lambda \gamma(t,1) - b(t))}\end{equation}
 which means that the optimal fraction $\hpi(t)$ to be placed in the risky asset is, using \eqref{38n}
 \begin{equation}\label{eq3.32}
 \hpi(t) = \frac{\hvar(t) S(t^-)}{X_{\hvar}(t)} = \frac{b(t)}{\gamma(t,1)(\lambda \gamma(t,1) - b(t))}.
 \end{equation}

\end{example}
\begin{remark}
To check that $\hvar$ is admissible, we have to verify that \eqref{eq2.A1} and \eqref{eq2.A2} hold for $\varphi =\hvar$. To this end, we see that condition \eqref{eq3.21a} suffices.
\end{remark}

\section{Robust duality}\label{sec5}
In this section we extend our study to a robust optimal portfolio problem and its dual. 

\subsection{Model uncertainty setup}

To get a representation of model uncertainty, we consider a family of probability measures $R = R^\kappa$    $\sim P$, with  Radon-Nikodym derivative on ${\cal F}_t$ given by 
\begin{equation*}
\frac{d(R^\kappa \mid {\cal F}_t)}{d(P \mid {\cal F}_t)} = Z^\kappa_t  
\end{equation*}
where, for $0 \leq t \leq T$, $Z^\kappa_t$ is a martingale of the form
\begin{equation*}
dZ^\kappa_t = Z^\kappa_{t^-} [\kappa_0(t)  dB_t + \int_\RB \kappa_1(t,\zeta) \tilde{N}(dt,d\zeta)] \;; \quad 
Z^\kappa_0 = 1.
\end{equation*}

Let $\mathbb K$  denote a  given set  of admissible 
scenario controls $\kappa= (\kappa_0, \kappa_1)$, $\FC_{t}$-predictable, 
 s.t. $\kappa_1(t,z) \geq -1 + \epsilon$, and
 $E [ \int_0^T \{ |\kappa_0^2(t) | + \int_\RB \kappa_1^2(t,z) \nu(dz)\} dt ] < \infty$.

By the Girsanov theorem, using the measure $R^\kappa$ instead of the original measure $P$  in the computations involving the price process $S(t)$, 
 is equivalent to using the original measure $P$ in the computations involving the \emph{perturbed price process} $S_\mu(t)$ instead of $P(t)$, where $S_\mu(t)$ is given by 
 \begin{equation}
\begin{cases}
\displaystyle dS_\mu(t)  =S_\mu(t^-)[ (b (t) + \mu (t))dt + \sigma (t) dB(t )
  + \int_\RB \gamma(t,\zeta) \tN (dt, d \zeta)]  \\
 S_\mu(0)  >0,
  \end{cases}
 \end{equation}
 with
 \begin{equation}
  \mu(t )= - \sigma(t )\kappa_0(t) - \int_\RB \gamma(t,\zeta) \kappa_1(t,\zeta) \nu(d\zeta)dt.
  \end{equation}
%
%
Accordingly, we now replace the price process $S(t)$ in \eqref{eq2.1} by the perturbed process
\begin{equation}\label{eq5.1}
\begin{cases}
\displaystyle dS_\mu(t)  = S_\mu(t^-)[ (b(t) + \mu(t))dt + \sigma(t) dB(t) 
  + \int_\RB \gamma(t,\zeta) \tN (dt, d \zeta)] \; ; \; 0 \leq t \leq T \\
 S_\mu(0)  >0,
 \end{cases}
 \end{equation}
 for some perturbation process $\mu(t)$, assumed to be predictable and satisfy
 $$E \left[ \int_0^T  |\mu(t)| dt\right] < \infty.$$
 Let $\mathbb M$ denote this set of perturbation processes $\mu$.
 Let $X = X_{\varphi,\mu}^x$ be  the corresponding wealth process given by
 \begin{equation}\label{eq5.4}
 \begin{cases}
 dX(t) \displaystyle = \varphi(t)S_\mu(t^-) [(b(t) + \mu(t))dt + \sigma(t) dB(t) + \int_\RB \gamma(t,\zeta) \tN(dt,d\zeta)] \; ; \; 0 \leq t \leq T \\
 X(0) = x > 0, 
 \end{cases}
 \end{equation}
 where $\varphi$ is an admissible portfolio, that is it belongs to the set  $\AC$ of $\FB$-predictable processes such that
  \begin{equation}
 \begin{cases}
\text{  \eqref{eq2.A1} and \eqref{eq2.A2} hold, }\\
 E \left[ \int_0^T \varphi(t)^2 S_\mu(t)^2 \left\{  (b(t) + \mu(t))^2 + \sigma^2(t) + \int_\RB \gamma^2(t,\zeta) \nu (d \zeta) \right\} dt \right] < \infty, \\
 X_{\varphi,\mu}(t) > 0   \text{ for all } t \in [0,T]  \text{ a.s. } \\
 \end{cases}
 \end{equation}
   for all $ \mu \in  \mathbb M.$ 
  
  \subsection{The robust primal and dual problems}\label{sec5.1}
 Let $\rho : \RB \rightarrow \RB$ be a convex penalty function, assumed to be $\CC^1$, and $U$  a utility function as in Section \ref{sec1}. We assume that $\rho(\mu)$ has a minimum at $\mu = 0$ and that $\rho(0) = 0$. Then $\rho(\mu)$ can be interpreted as a penalization for choosing $\mu \neq 0$.
 \begin{definition}\label{defi5.1}
 The {\it robust primal problem} is, for given $x>0$, to find $(\hvar, \hmu) \in \AC \times \mathbb M $ such that
\begin{equation}\label{eq5.5}
\inf_{\mu \in \mathbb M} \sup_{\varphi \in \AC}   I(\varphi, \mu)  = I(\hvar, \hmu) = \sup_{\varphi \in \AC} \inf_{\mu \in \mathbb M} I(\varphi, \mu),
\end{equation}
where
\begin{equation}\label{eq5.6}
I(\varphi, \mu) = E \left[ U(X_{\varphi, \mu}^x(T)) + \int_0^T \rho(\mu(t))dt \right].
\end{equation}
\end{definition}
The problem \eqref{eq5.5} is a stochastic differential game. To handle this, we use an extension of the maximum principle to games, as presented in, e.g., \cite{OS4}. We 
 obtain the following characterization of a solution (saddle point) of \eqref{eq5.5}:
\begin{proposition}[Robust primal problem and associated constrained FBSDE]\label{theo5.2}
 A pair $(\hvar, \hmu) \in \AC \times \mathbb M$ is a solution of the robust primal problem \eqref{eq5.5} if and only if the solution $X(t)$, $(p_1,q_1, r_1)$ of the FBSDE consisting of the SDE \eqref{eq5.4} and the BSDE
 \begin{equation}\label{eq5.11}
 \begin{cases}
 \displaystyle dp_1(t) = q_1(t) dB(t) + \int_\RB r_1(t,\zeta) \tN (dt, d\zeta) \; ; \; 0 \leq t \leq T \\
 p_1(T) = U'(X_{\hvar,\hmu}^x(T))
 \end{cases}
 \end{equation}
 satisfies 
 \begin{equation}\label{eq5.9}
(b(t) + \hmu(t)) p_1(t) + \sigma(t) q_1(t) + \int_\RB \gamma(t,\zeta) r_1(t,\zeta) \nu (d\zeta) = 0 \; ; \; t \in [0,T]
\end{equation}
\begin{equation}\label{eq5.10}
\rho'(\hmu(t)) + \hvar(t) S_{\hmu}(t^-) p_1(t) = 0 \; ; \; t \in [0,T].
\end{equation}
 \end{proposition}
\dproof
Define the Hamiltonian by
\begin{equation}\label{eq5.7}
H_1(t,x,\varphi,\mu,p,q,r) = \rho(\mu) + \varphi  S_\mu(t^-) \left[ (b(t) + \mu)p + \sigma(t)q + \int_\RB \gamma(t,\zeta) r(\zeta)\nu(d\zeta)\right].
\end{equation}
The associated BSDE for the adjoint processes $(p_1,q_1,r_1)$ is \eqref{eq5.11}.

The first order conditions for a maximum point $\hvar$ and a minimum point $\hmu$, respectively, for the Hamiltonian are
given by \eqref{eq5.9} and \eqref{eq5.10}. 
Since $H_1$ is concave with respect to $\varphi$ and convex with respect to $\mu$, these first order conditions are also {\it sufficient} for $\hvar$ and $\hmu$ to be a maximum point and a minimum point, respectively.  
 \fproof
 
  
%
We now study a dual formulation of the robust primal problem \eqref{eq5.5}. 
Let now  $\MC$ be the family of positive measures $Q = Q_{\theta, \mu}$ of the form
\begin{equation}\label{eq2.5a}
dQ_{\theta, \mu}(\omega) = G_{\theta,\mu}(T) dP(\omega) \text{ on } \FC_T,
\end{equation}
where
 $G(t) = G_{\theta,\mu}^y(t)$ is given by
 \begin{equation}\label{eq5.14}
 \begin{cases}
\displaystyle  dG(t) = G(t^-) \left[ \theta_0(t) dB(t) + \int_\RB \theta_1(t,\zeta) \tN(dt,d\zeta)\right] \; ; \; 0 \leq t \leq T \\
G(0) = y > 0
\end{cases}
\end{equation}
and $(\theta, \mu)$ is such that $\mu \in \mathbb M$ and $\theta= (\theta_0, \theta_1)$ is a  predictable processes  satisfying \eqref{eq2.7} and 
\begin{equation}\label{eq5.15}
b(t) + \mu(t) + \sigma(t)\theta_0(t) + \int_\RB \gamma(t,\zeta) \theta_1(t,\zeta) \nu(d\zeta) = 0 \; ; \; t \in [0,T].
\end{equation}
 We let $ \Lambda$ denote the set of  such  processes $(\theta, \mu)$.
If $y=1$, then the measure $Q_{\theta, \mu}$ 
is an ELMM for the perturbed price process $S_\mu$ in \eqref{eq5.1}. 
 \begin{definition}\label{defi5.3}
 The {\it robust dual problem} is for given $y >0$,  to find $( \tth,\tilde \mu)  \in \Lambda$ such that
 \begin{equation}\label{eq5.12}
 \sup_{(\theta, \mu) \in  \Lambda} J(\theta,\mu) = J(\tth, \tilde \mu) 
 \end{equation}
 where
 \begin{equation}\label{eq5.13}
 J(\theta,\mu) = E \left[ - V(G_{\theta, \mu}^y(T)) - \int_0^T \rho (\mu(t))dt \right],
 \end{equation}
 and $V$ is the convex conjugate function of $U$, as in Section \ref{sec1}.
 \end{definition}
 \begin{proposition}[Robust dual problem and its associated constrained FBSDE.]\label{th5.4}
 A pair $(\tth, \tmu) \in \Lambda$ is a solution of the robust dual problem \eqref{eq5.12}-\eqref{eq5.13} if and only the solution $G(t)$, $(p_2, q_2, r_2)$ of the FBSDE consisting of the FSDE \eqref{eq5.14} and the BSDE
  \begin{equation}\label{eq5.22}
  \begin{cases}
 \displaystyle   dp_2(t) = K(q_2,r_2)(t) [b(t) + \tmu(t)]dt  + q_2(t) dB(t) + \int_\RB r_2(t,\zeta) \tN(dt,d\zeta) \; ; \; t \in [0,T] \\
  p_2(T) = - V'(G_{\tth,\tmu}^y(T))
  \end{cases}
  \end{equation} 
  with $K(q,r)(t)$ defined as in \eqref{eq2.19}, 
  satisfies the two equations
  \begin{equation} \label{eq5.21}
  G_{\tth,\tmu}^y (t) q_2(t) + \rho'(\tmu(t)) \sigma(t)  = 0, 
  \end{equation}
  \begin{equation}
  \label{eq5.20}
  G_{\tth,\tmu}^y (t) r_2(t, \zeta) + \rho'(\tmu(t)) \gamma(t, \zeta) = 0.
  \end{equation}
  
%
 \end{proposition}
 
 \dproof
We proceed as in the proof of Proposition \ref{prop2.2}:
The Hamiltonian for the constrained stochastic control problem \eqref{eq5.12} is
 \begin{align}\label{eq5.21a}
 & H_2^L(t,g,\theta_0, \theta_1,\mu,p,q,r)\nonumber\\
 &:= -\rho(\mu)+g \theta_0 q + g \int_\RB \theta_1(\zeta) r(\zeta) \nu(d\zeta)
    + L(t) \left( b(t) +\mu(t) + \sigma(t) \theta_0 + \int_\RB \gamma(t,\zeta) \theta_1(\zeta) \nu (d\zeta)\right),
   \end{align}
   where $L(t)$ is the Lagrange multiplier process.
   
  The first order conditions for a maximum point $(\tth, \tmu)$ for $H_2^L$ are  $\nabla_{\theta} H_2^L = 0$ and 
   $ \left( \frac{\partial H_2^L}{\partial \mu} \right)= 0$ 
  which reduce to 
  \eqref{eq5.21}-\eqref{eq5.20}.
  Then, as in \eqref{eq2.38} we see that the corresponding BSDE for the adjoint processes $(p_2,q_2,r_2)$ is given by \eqref{eq5.22}.\\
  
  Since $H_2$ is concave w.r.t. $\mu$ and $\theta$, these necessary optimality 
  conditions are also sufficient. 
     \fproof
     
 \subsection{Relations between robust primal and robust dual problems} \label{sec5.3}
 We now use the characterizations above of the solutions $(\hvar, \hmu )\in \AC \times \mathbb M$ and $(\tth, \tmu) \in \Lambda$ of the robust primal and the robust dual problem, respectively, to find the relations between them.
\begin{theorem} \label{th3.5} {\bf (i) From robust primal to robust dual.}

    Assume  $(\hvar, \hmu) \in \AC \times \mathbb M$ is a solution of the robust primal problem and let $(p_1, q_1, r_1)$ be  the associated adjoint processes  solution of the FBSDE \eqref{eq5.4} \& \eqref{eq5.11} and satisfying \eqref{eq5.9}-\eqref{eq5.10}.
    Define 
 \begin{align}
& \tmu := \hmu \label{eq5.23} \\
& \tth_0(t) := \frac{q_1(t)}{p_1(t)} \text{ ; } \; \tth_1(t,\zeta) = \frac{r_1(t,\zeta)}{p_1(t)} \label{eq5.24}
 \end{align}
 and suppose they satisfy \eqref{eq2.7}.
 Then, they are  optimal for the dual problem with initial value $y = p_1(0).$
 Moreover  
 \begin{equation}\label{eq5.27}
p_1(t) = G_{\tth, \tmu} (t) \; ; \; t \in [0,T].
\end{equation}
In particular,
\begin{equation}\label{eq5.28}
U'(X_{\hvar, \hmu}(T)) = G_{\tth, \tmu} (T).
\end{equation}

{\bf (ii) From robust dual to robust primal}
 Let  $(\tth, \tmu) \in \Lambda$ be  optimal for the  robust dual problem  \eqref{eq5.12}-\eqref{eq5.13} and let $(p_2, q_2, r_2)$ be the associated adjoint processes satisfying \eqref{eq5.22} with the constraints 
 \eqref{eq5.20} and  \eqref{eq5.21}.
 Define
 \begin{align}
& \hmu := \tmu \label{eq5.31} \\
& \hvar(t ):= \frac{K(q_2,r_2)(t)}{ S_{\hmu}(t^-)} \; ; \; t \in [0,T].  \label{eq5.32}
 \end{align}
 Assume that $\hvar \in \AC$. 
 Then $(\hmu,  \hvar)$  are optimal for primal problem with initial value $x = p_2(0).$
Moreover, 
\begin{equation}\label{eq4.42a}
 p_2(t) = X_{\hvar,\hmu}(t) \quad t \in [0,T].
 \end{equation}
 In particular
\begin{equation}\label{eq4.42}
- V'( G_{\tilde{\theta}}(T)) = X_{\hvar,\hmu}(T).
\end{equation}

\end{theorem}
   
\dproof
{\bf (i) }
 Let  $(\hvar, \hmu) \in \AC \times \mathbb M$ is a solution of the robust primal problem and let $(p_1, q_1, r_1)$ be as in Proposition
 \ref{theo5.2}, i.e. assume that $(p_1, q_1, r_1)$ solves the FBSDE \eqref{eq5.4} and \eqref{eq5.11} and satisfies \eqref{eq5.9}-\eqref{eq5.10}.
 
 We want to find the solution $(\tth, \tmu) \in \Lambda$ of the robust dual problem. By Proposition \ref{th5.4} this means that we must find a solution $(p_2, q_2, r_2)$ of the FBSDE \eqref{eq5.14} and \eqref{eq5.22} which satisfies \eqref{eq5.20}-\eqref{eq5.21}. 
 To this end, choose $\tmu,  \tth_0,  \tth_1$
given  in \eqref{eq5.23}-\eqref{eq5.24}.
 Then by \eqref{eq5.9} we have
 \begin{equation}\label{eq5.25}
 b(t) + \tmu(t)+ \sigma(t) \tth_0(t) + \int_\RB \gamma(t,\zeta) \tth_1(t,\zeta) \nu (d\zeta) = 0.
 \end{equation}
 Assume that \eqref{eq2.7} holds. 
 Then $(\tmu,\tth) \in  \Lambda$.
 Substituting \eqref{eq5.24} into \eqref{eq5.11}, we obtain
 \begin{equation}\label{eq5.26}
 \begin{cases}
\displaystyle  dp_1(t) = p_1(t^-) \left[ \tth_0(t) dB(t) + \int_\RB \tth_1(t,\zeta) \tN (dt, d\zeta) \right] \; ; \; t \in [0,T] \\
 p_1(T) = U'(X_{\hvar, \hmu}(T)).
 \end{cases}
 \end{equation}
 Comparing with \eqref{eq5.14} we see that
 $$\frac{dG_{\tth, \tmu} (t)}{G_{\tth, \tmu} (t)} = \frac{dp_1(t)}{p_1(t)}$$
 and hence, for  $y = G_{\tth, \tmu} (0)  =p_1(0)> 0$ we get 
\eqref{eq5.27} and \eqref{eq5.28}.
Define
\begin{equation}\label{eq5.29}
p_2(t) := X_{\hvar, \hmu}(t), q_2(t) := \hvar(t) \sigma(t) S_{\hmu}(t^-), r_2(t, \zeta) := \hvar(t) \gamma(t,\zeta)S_{\hmu}(t^-).
\end{equation}
Then by \eqref{eq5.4} and \eqref{eq5.28}, combined with \eqref{eq1.7},
\begin{equation}\label{eq5.30}
\begin{cases}
\displaystyle dp_2(t) = \hvar(t) S_{\hmu}(t^-) \left[(b(t) + \hmu(t)) dt + \sigma(t) dB(t) + \int_\RB \gamma(t,\zeta) \tN (dt, d\zeta) \right] \\
\displaystyle \quad = K(q_2,r_2)(t) [b(t) + \hmu(t)] dt + q_2(t) dB(t)
 + \int_\RB r_2(t,\zeta) \tN (dt, d\zeta) \; ; \; 0 \leq t \leq T \\
p_2(T) = X_{\hvar, \hmu}(T) = - V'(G_{\tth, \tmu} (T)).
\end{cases}
\end{equation}
Hence $(p_2, q_2, r_2)$ solves the BSDE \eqref{eq5.22}, as requested.
It remains to verify that \eqref{eq5.20} and \eqref{eq5.21} hold:
By \eqref{eq5.29} we have
$$- q_2(t) \gamma(t, \zeta) + \sigma(t) r_2(t,\zeta) = \sigma(t)[- \hvar(t) S_{\hmu}(t^-)\gamma(t, \zeta) + \hvar(t) S_{\hmu}(t^-)\gamma(t,\zeta)] = 0,$$
which is \eqref{eq5.20}.
By \eqref{eq5.23}, \eqref{eq5.27}, \eqref{eq5.29} and \eqref{eq5.10},
$$\rho'(\tmu) + G_{\tth,\tmu}(t) q_2(t) = \rho'(\hmu) + p_1(t) \hvar(t) \sigma(t) S_{\hmu}(t^-)= 0,$$
which is \eqref{eq5.21}.


{\bf (ii)}
Next, assume that $(\tth, \tmu) \in \Lambda$ is optimal for the robust dual problem \eqref{eq5.12}-\eqref{eq5.13} and let $(p_2, q_2, r_2)$ be as in Proposition \ref{th5.4}.
We will find $(\hvar, \hmu) \in \AC \times \mathbb M$ and $(p_1, q_1, r_1)$ satisfying Proposition \ref{theo5.2}. Choose
$\hmu$ and 
$\hvar$ given in \eqref{eq5.31}--\eqref{eq5.32}   and assume that $\hvar$ is admissible.
Then by \eqref{eq5.22} and \eqref{eq5.20}
$$\begin{cases}
\displaystyle dp_2(t) = \hvar (t) S_{\hmu}(t^-) \left[ (b(t) + \hmu(t) \sigma(t)) dt + \sigma(t) dB(t) + \int_\RB \gamma(t,\zeta) \tN (dt, d\zeta) \right] \; ; \; 0 \leq t \leq T \\
p_2(T) = - V'(G_{\tth,\tmu}(T)).
\end{cases}
$$
Hence, with  $x = p_2(0) >0$, \eqref{eq4.42a}  holds. 
In particular
\begin{equation}\label{eq5.34}
X_{\hvar, \hmu}(T) = p_2(T) = - V'(G_{\tth,\tmu} (T)), \text{ i.e. } G_{\tth,\tmu} (T) = U' (X_{\hvar, \hmu}(T)).
\end{equation}
We now verify that  with $\varphi = \hvar, \mu = \hmu$, and $p_1, q_1, r_1$ defined by 
\begin{equation}\label{eq5.35}
p_1(t) := G_{\tth,\tmu} (t),  \; q_1(t) := G_{\tth,\tmu} (t) \tth_0(t),  \; r_1(t,\zeta) := G_{\tth,\tmu} (t) \tth_1(t,\zeta), 
\end{equation}
all the conditions of Proposition \ref{theo5.2} hold: 
By \eqref{eq5.14} and \eqref{eq5.34},
\begin{equation}\label{eq5.36}
\begin{cases}
\displaystyle dp_1(t) = dG_{\tth, \tmu} (t) = G_{\tth, \tmu} (t^-) \big( \tth_0(t) dB(t) + \int_\RB \tth_1(t,\zeta) \tN(dt,d\zeta) \big)\\
= \displaystyle q_1(t)dB(t) + \int_\RB r_1(t,\zeta) \tN(dt,d\zeta)
 \quad  \; ; \; 0 \leq t \leq T \\
p_1(T) = G_{\tth,\tmu} (T) = U'(X_{\hvar, \hmu}(T)).
\end{cases}
\end{equation}
Hence \eqref{eq5.11} holds.
It remains to verify \eqref{eq5.9} and \eqref{eq5.10}. By \eqref{eq5.35} and \eqref{eq5.15} for $\theta = \tth$, we get
\begin{align*}
(b(t) + \hmu(t)& )p_1(t) + \sigma(t) q_1(t) + \int_\RB \gamma(t,\zeta) r_1 (t,\zeta) \nu (d\zeta) \\
&= G_{\tth,\tmu} (t) \left[ b(t) + \hmu(t)  + \sigma(t) \tth_0(t) + \int_\RB \gamma(t,\zeta) \tth_1(t,\zeta) \nu(d\zeta) \right] = 0,
\end{align*}
which is \eqref{eq5.9}.
By \eqref{eq5.31}, \eqref{eq5.32}, \eqref{eq5.35} and \eqref{eq5.21} we get
$$\rho'(\hmu(t)) + \hvar(t) S_{\hmu}(t^-) \sigma(t) p_1(t) = \rho'(\tmu(t)) + q_2(t) G_{\tth,\tmu} (t) = 0,$$
which is \eqref{eq5.10}. 
\fproof
\subsection{Illustrating examples}
\begin{example}\label{ex2}
\rm
 We consider a robust version of the classical Merton type optimal portfolio problem:
 We assume that there exists a constant $C > 0$ such that
\begin{equation}
\frac{|b(t)|}{|\sigma(t)|} \leq C \; ; \; 0 \leq t \leq T. 
\end{equation}
 We want to study 
 \begin{equation}\label{eq5.5b}
\inf_{\mu \in  \mathbb M} \sup_{\varphi \in \AC} E \left[ U(X_{\varphi, \mu}(T)) + \int_0^T \rho(\mu(t))dt \right]
\end{equation}
in the case with no jumps ($N = \gamma = 0, \sigma \neq 0$).
Then there is only one ELMM for the price process $S_\mu(t)$ for each given $\mu(t)$.
 So  $\theta = \theta_0 = - \frac{b(t)+\mu(t)}{\sigma(t)}$     
  and  the corresponding robust dual problem simplifies to 
 \begin{equation}\label{peq5.12}
 \sup_{\mu \in  \mathbb M}  E \left[ - V(G_\mu(T)) - \int_0^T \rho (\mu(t))dt \right],
 \end{equation}
 where 
  \begin{equation}\label{6.3}
\displaystyle  dG_{\mu}(t) = - G_\mu(t^-)  \frac{b(t)+\mu(t)}{\sigma(t)} dB_t  \; ; \; 0 \leq t \leq T;  \; \; 
G_\mu(0) = y > 0.
\end{equation}
  The first order conditions for the Hamiltonian reduce to:
  \begin{equation}\label{equ4.44}
\tmu(t) =({ \rho'})^{-1} (-  \frac{G_{\tmu} (t) q_2(t)}{\sigma(t)}) 
  \end{equation}
which substituted into the adjoint BSDE equation gives:
 \begin{equation}\label{6.5}
  \begin{cases}
 \displaystyle dp_2(t )=  \frac{q_2(t)}{\sigma(t)}
 [  b(t) + 
 ({ \rho'})^{-1} (-  \frac{G_{\tmu} (t) q_2(t)}{\sigma(t)}) 
  ] dt + q_2(t )dB_t; 
\; ; \; t \in [0,T] \\
  p_2(T) = - V'(G_{\tmu}(T)).
  \end{cases}
  \end{equation}
 We get that 
 $\tmu$ is optimal for the robust dual problem if and only if there is a solution $(p_2,q_2,G_{\tmu})$ of the FBSDE consisting of \eqref{6.5} and 
  \eqref{6.3} with the constraint \eqref{equ4.44}. 
 Hence,  by Theorem \ref{th3.5}(ii), the optimal $\hmu$ for the primal robust problem is given by $\hmu:=\tmu$, and the optimal 
 portfolio is
 \begin{equation} \label{eq5.32bis}
\hvar(t) =  \frac{K(q_2,r_2)(t)}{ S_{\hmu}(t^-)} = \frac{q_2(t)}{\sigma(t) S_{\tmu}(t^-)} \; ; \; t \in [0,T].
\end{equation}
Now assume that 
 \begin{equation}\label{ln}
U(x) = \ln x
\;\;  \text{ and } \; 
\rho(x) = \frac{1}{2} x^2.
\end{equation}
Then $V(y) = - \ln y - 1.$ \\
If $b(t)$  and $\sigma(t)$ are deterministic,  we can solve \eqref{peq5.12} by dynamic programming, and we get
\begin{equation}\label{eq450}
\tmu(t) =  - \frac{b(t)}{2} ; \; t \in [0, T].
\end{equation}
In view of this, it is natural to guess that \eqref{eq450} is the optimal choice of $\mu$ also when $b(t)$  and 
$\sigma(t)$ are ${\cal F}_t$-adapted processes. To verify this we  have to show that the system  \eqref{6.3}-\eqref{6.5} is consistent. 
This system is now the following 
\begin{align}\label{55}
G_{\tmu}(t)&= y \exp\left( - \int_0^t \frac{b(s)}{2\sigma(s)} dB(s) - \frac{1}{2} (\frac{b(s)}{2\sigma(s)})^2 ds \right)\\
q_2(t) & =  \frac{1}{G_{\tmu}(t)} .  \frac{b(t)}{2\sigma(t)} \label{451} \\
dp_2(t) &= \frac{1}{G_{\tmu}(t)} \left[  \frac{b(t)}{2\sigma(t)} dB(t) + (\frac{b(t)}{2\sigma(t)})^2 dt \right] \;  ; \;
p_2(T) = \frac{1}{G_{\tmu}(T)} \label{452}
\end{align}

which gives
\begin{equation}
 \frac{1}{G_{\tmu}(t)} =  \frac{1}{y} \exp(  \int_0^t \frac{b(s)}{2\sigma(s)} dB(s) + \frac{1}{2} (\frac{b(s)}{2\sigma(s)})^2 ds)
\end{equation}
i.e.
\begin{equation}\label{57}
d ( \frac{1}{G_{\tmu}(t)} ) =  \frac{1}{G_{\tmu}(t)}  \left[ \frac{b(t)}{2\sigma(t)} dB(t) + (\frac{b(t)}{2\sigma(t)})^2 dt \right] .
\end{equation}
We see that \eqref{452}  is in agreement with \eqref{57} with $p_2(t) = \frac{1}{G_{\tmu}(t)}$, and this proves that 
$\tmu(t)$  given by \eqref{eq450}
is indeed optimal also when $b$ and $\sigma$ are stochastic. 
The corresponding optimal portfolio for the robust utility maximization  problem with initial value $x = \frac{1}{y} $, is,  by \eqref{eq5.32}, 
\begin{equation}\label{eq455}
\hvar(t) = \frac{b(t)}{ G_{\tmu}(t) 2\sigma^2(t) S_{\tmu}(t)} ; \;\; t \in [0, T].
\end{equation}
which means that the optimal fraction of wealth to be placed in the risky asset is
\begin{equation}\label{eq4.54}
\hpi(t )= \frac{\hvar(t)S_{\tmu}(t^-)}{\hat{X}(t)} = \frac{b(t)}{ 2 \sigma^2(t)}
\end{equation}
We have thus proved:
\begin{proposition}
Suppose \eqref{ln}
  holds. Then the optimal scenario $\hmu=\tmu$ and optimal portfolio $\hvar$ for the robust primal problem 
\eqref{eq5.5b} are given by \eqref{eq450} and \eqref{eq455}, 
respectively, with $G_{\tmu}(t)$ as in \eqref{55}.
\end{proposition}
\begin{remark}\label{rem4.8}
Comparing \eqref{eq4.54} with the solution of the Merton problem in the classical, non-robust case,
we see that the optimal fraction to be placed in the risky asset in the robust case is just half of the optimal fraction in the non-robust case.
\end{remark}
\end{example}

\begin{example} \rm 
We consider a robust version of Example \ref{examples}. In this case the perturbed price $S=S_\mu$  
is \begin{equation}
dS(t) = S(t^-)[(b(t) + \mu(t)) dt + \gamma (t,1) d \tN(t)] \; ; \; S(0) > 0
\end{equation}
and  the wealth process $X = X_{\varphi,\mu}^x$ associated to a portfolio $\varphi$ is 
\begin{equation}
dX(t) = \varphi(t) S(t^-) [b(t) dt + \gamma(t,1) d \tN(t)] \; ; \; X(0) = x > 0.
\end{equation}
We again choose the logarithmic utility function $U(x) = \ln(x)$ and the quadratic penalty function $\rho(x) = \frac{1}{2} x^2$. 

Thus the robust  primal  problem is 
to find $(\hvar, \hmu) \in \AC \times \mathbb M $ such that
\begin{equation}\label{eq5.5}
\inf_{\mu \in \mathbb M} \sup_{\varphi \in \AC}   
E \left[  \ln X_{\varphi, \mu}^x(T)) + \int_0^T \mu^2(t)dt \right]
= E \left[  \ln X_{\hvar, \hmu}^x(T)) + \int_0^T \hmu^2(t) dt \right].
\end{equation}
The corresponding dual problem is 
  to find $( \tth,\tilde \mu)  \in \Lambda$ such that
 \begin{equation}\label{eq5.12}
 \sup_{(\theta, \mu) \in  \Lambda} 
 E \left[ \ln (G_{\theta, \mu}^y(T)) - \int_0^T  \mu^2(t)dt \right]
= E \left[ \ln (G_{\tth, \tilde \mu}^y(T)) - \int_0^T  \tilde \mu^2(t)dt \right].
 \end{equation}
First note from \eqref{eq5.15}  that for each $\mu$ there is only  one admissible element 
process $\theta$ given by
\begin{equation}
\theta_1(t,1) = \tth_1(t,1) = - \frac{b(t) + \mu(t)}{\lambda \gamma (t,1)}.
\end{equation}
Assume that (see \eqref{eq2.7})
\begin{equation}\label{cond}
\frac{b(t) + \mu(t)}{\lambda \gamma(t,1)} < 1 ; \; t \in [0, T] . 
\end{equation}
Then we get
\begin{align}\label{eq3.22p}
G^y_{\tth_1, \tilde \mu}(t) & = y \exp \big( \int_0^t \ln (1 - \frac{b(s) + \tilde \mu(s)}{\lambda \gamma(s,1)}) d \tN(s) \nonumber \\
 & + \lambda \int_0^t \{ \ln ( 1 - \frac{b(s)+ \tilde \mu(s)}{\lambda \gamma(s,1)}) + \frac{b(s)+ \tilde \mu(s)}{\lambda \gamma (s,1)} \} ds \big) \; ; \; 0 \leq t \leq T.
 \end{align}
 In this case, $K(q_2, r_2)(t) = \frac{r_2(t,1)}{\gamma (t,1)}$ and the BSDE \eqref{eq5.22} becomes: 
 \begin{equation}\label{eq5.22p}
  \begin{cases}
 \displaystyle   dp_2(t) =  \frac{ r_2(t,1)}{\gamma (t,1)}[b(t) + \tmu(t)]dt  + \int_\RB r_2(t,1) d\tN(t) \; ; \; t \in [0,T] \\
  p_2(T) = \frac{1}{G_{\tth,\tmu}^y(T)}.
  \end{cases}
  \end{equation} 
 To solve this equation, we proceed as in Example \ref{examples}.  We then get:
 $$\hp_2 = \frac{1}{G_{\tth,\tmu}^y}\; ; \; \hr_2 = \hp_2 \psi$$
 with $\psi$ given by
 \begin{equation}
 \psi(t) = \frac{b(t) + \tmu(t)}{\lambda \gamma(t,1) - ( b(t) + \tmu(t))}, \; t \in [0,T].
 \end{equation}
From \eqref{eq5.20}, we get the equation: 
\begin{equation}
\tmu(t) = - \frac{G_{\tth,\tmu}^y(t)   \hp_2 (t)    \psi(t)}  {\gamma(t,1)} = 
 - \frac{b(t)  + \tmu(t)  }{\gamma(t,1) (\lambda \gamma(t,1) - (b(t)  + \tmu(t)))}, 
\end{equation} i.e.
\begin{equation}
\gamma(t,1) \tmu^2(t) + (\gamma(t,1) b(t) - \lambda \gamma^2(t,1) - 1) \tmu(t) - b(t) = 0.
\end{equation}
The root of this quadratic equation which satisfies \eqref{cond} is 
\begin{equation}\label{eq3.65}
\tmu(t) = \frac{1}{2 \gamma(t,1)} ( - \gamma(t,1) b(t) + \lambda \gamma^2(t,1) + 1 
 - \sqrt{\Delta}).
\end{equation}
with 
\begin{align*} 
\Delta & = (\gamma(t,1) b(t) -  \lambda \gamma^2(t,1) - 1)^2 + 4 \gamma(t,1) b(t) \\ 
& = (\gamma(t,1) b(t) -  \lambda \gamma^2(t,1) + 1)^2 + 4 \lambda \gamma^2(t,1).
\end{align*} 
From Theorem \ref{th3.5} we conclude that the solution of the robust primal problem \eqref{eq5.5} 
is $\hat \mu(t) = \tmu(t)$   given by \eqref{eq3.65}, and
$$\hat \varphi(t) = \frac{\hr_2 (t)}{\gamma(t,1) S(t^-)} = \frac{\hp_2(t) \psi(t) }{\gamma(t,1) S(t^-)} 
= \frac{X_{\hat \varphi, \hat\mu}^x(t) \psi(t) }{\gamma(t,1) S(t^-)} $$
with $x = \frac{1}{y}$.
The optimal fraction of the wealth invested in the risky asset is
$$ \hat \pi(t) = \frac{\hat \varphi(t)  S(t^-)}{X_{\hat \varphi, \hat\mu}^x(t) } = 
\frac{\psi(t)}{\gamma(t,1)} = \frac{ b(t) + \hat \mu(t)}{\gamma(t,1)(\lambda \gamma(t,1) 
- (b(t) + \hat \mu(t)))} = - \hat \mu(t). $$

We summarize this as follows: 
\begin{proposition} 
The optimal pair $(\hat \mu, \hat \varphi) \in \AC \times \mathbb M$   for Problem \ref{eq5.5}
is given by  $   \displaystyle \hat \varphi (t) =    \frac{\hat \pi (t) X_{\hat \varphi, \tmu}^x(t) }{S(t^-)}$ with 
$\hat \pi (t) = - \tmu(t)$ and $\tmu(t)$ given by \eqref{eq3.65}.

\end{proposition}

\end{example}
\appendix
\section{Maximum principles for optimal control}
Consider the following controlled stochastic differential equation
\begin{align}\label{1}
dX(t) &= b(t, X(t), u(t), \omega) dt + \sigma(t, X(t), u(t), \omega)  dB(t) \\ & + \int_\RB \gamma(t,   X(t), u(t), \omega, \zeta)
\tN (dt, d\zeta) \; ; \; 0 \leq t \leq T \;  ; \; 
X(0) = x \in \RB.  \nonumber 
\end{align}
The performance functional is given by
\begin{equation}
J(u) = E \left[ \int_0^T f(t, X(t), u(t), \omega) dt + \phi(X(T), \omega) \right]
\end{equation}
where $T>0$ and $u$ is in a given family $\AC$ of admissible $\FC$-predictable controls.
 For $u \in \AC$ we let $X^u(t)$ be the solution of \eqref{1}. We assume this solution exists, is unique  and satisfies
 \begin{equation}
 E[ \int_0^T  |X^u(t) |^{2 
 }
  dt ] < \infty.
 \end{equation} 
We want to find $u^* \in \AC$ such that 
\begin{equation}\label{pb}
\sup_{u \in \AC } J(u)  = J(u^*).
\end{equation}
We make the following assumptions
\begin{align}
f  \in C^1 \text { and } E[\int_0^T  | \nabla f |^2(t) dt ] < \infty, \\
b, \sigma, \gamma \in C^1 \text { and } E[\int_0^T ( | \nabla b |^2  +   | \nabla \sigma |^2   + 
 \| \nabla \gamma \|^2  ) (t) dt ] < \infty, \\
 \text{ where }  \| \nabla \gamma(t, \cdot) \|^2 := \int_\RB \gamma^2(t, \zeta) \nu (d \zeta) \nonumber \\
 \phi \in C^1 \text { and } \text{ for all } u \in \AC,  \;  E[ \phi'(X(T))^{2 }] < \infty.
 \end{align}
 Let $\UB$ be a convex closed set containing all possible control values $u(t); t \in [0,T] $. 
 
 The Hamiltonian associated to the problem \eqref{pb} is defined by
 $$H: [0,T] \times  \RB \times \UB \times \RB  \times \RB \times {\cal R } \times \Omega \mapsto \RB$$
 $$
 H(t,x, u,p,q,r, \omega)  = f(t,x,u, \omega) + b (t,x,u, \omega) p + \sigma(t,x,u, \omega) q + \int_\RB \gamma(t,x,u, \zeta, 
 \omega) r(t, \zeta) \nu(d \zeta).$$
 For simplicity of notation the dependence on $\omega$ is suppressed in the following.
 We assume that $H$ is Fr\'echet differentiable in the variables $x,u$. 
 We let $m$ denote the Lebesgue measure on $[0,T]$.

 The associated BSDE for the adjoint processes $(p,q,r)$ is
 \begin{equation}\label{A8}
\begin{cases}
\displaystyle dp(t) = - \frac{\partial{H}}{\partial{x}}(t) + q(t) dB(t) + \int_\RB r(t,\zeta) \tN (dt,d\zeta) \; ; \; 0 \leq t \leq T \\
 p(T) =  \phi'(X(T)).
 \end{cases}
 \end{equation}
 Here and in the following we are using the abbreviated notation
 $$\frac{\partial H}{\partial x} (t) = \frac{\partial H}{\partial x} (t, X(t), u(t)) \text{ etc } $$
 We first formulate a sufficient maximum principle. 
 \begin{theorem}[Sufficient maximum principle]\label{thA1}
 Let $\hu \in \AC$  with corresponding solutions $\hat{X}$, $\hp, \hq, \hr$ of equations \eqref{1}-\eqref{A8}.
 Assume the following:
 \begin{itemize}
 \item  The  function $x \mapsto \phi(x)$ is concave 
 \item (The Arrow condition) The function
 \begin{equation}
 {\cal H }(x) :=  \sup_{v \in \UB} H(t, x, v, \hp(t),\hq(t),\hr(t, \cdot))  
 \end{equation}
 is concave for all $t \in [0,T]$.
 \item 
 \begin{equation}
  \sup_{v \in \UB} H(t, \hX(t), v, \hp(t),\hq(t),\hr(t, \cdot)) = H(t, \hX(t), \hu(t), \hp(t),\hq(t),\hr(t, \cdot)) ; \; t \in [0,T].
 \end{equation}
 \end{itemize}
 Then $\hu$ is an optimal control for the problem \eqref{pb}.
  \end{theorem}  
Next,  we state a necessary maximum principle.  For this,  we need the following assumptions: 
 \begin{itemize}
 \item For all $t_0 \in [0, T] $ and all bounded $\FC_{t_0}$-measurable random variables $\alpha(\omega)$ 
 the control 
 $$ \beta(t) := \chi_{[t_0, T]}(t) \alpha(\omega)$$ belongs to $\AC$.

 \item For all $u, \beta \in \AC$ with $\beta$ bounded, there exists $\delta >0$ such that the control 
 $$\tilde{u}(t) := u(t) + a \beta(t) ; \; t \in [0,T]$$
 belongs to $\AC$ for all $a \in ( - \delta, \delta)$.

 \item The derivative process
 $$x(t) := \frac{d}{da} X^{u + a \beta} (t) \mid_{a =0},$$
 exists and belongs to $L^2(dm \times dP)$, and 
 \begin{equation}\label{A23}
\begin{cases}
\displaystyle 
 dx(t) = \{  \frac{\partial b}{\partial x} (t) x(t) +  \frac{\partial b}{\partial u} (t) \beta(t) \} dt + 
  \{
  \frac{\partial \sigma}{\partial x} (t) x(t)   +   \frac{\partial \sigma}{\partial u} (t) \beta(t)   \} dB(t)  \\ \qquad 
\displaystyle +   \int_\RB   \{ \frac{\partial \gamma}{\partial x} (t,  \zeta) x(t) + \frac{\partial \gamma}{\partial u} (t,  \zeta) \beta(t)   \} 
    \tilde{N}(dt, d \zeta) \\
 x(0) =0 
 \end{cases}
 \end{equation}
  \end{itemize}

  \begin{theorem}[Necessary maximum principle]\label{thmA.2}
  The following are equivalent
  \begin{align*}
  \bullet  & \;\; \frac{d}{da} J(u + a \beta) \mid_{a=0} = 0 \text{ for all bounded } \beta \in \AC \\ 
    \bullet & \;\;  \frac{\partial H}{\partial u}(t)  = 0 \text{ for all }  t \in [0, T].
   \end{align*}
   \end{theorem}
   For proofs of these results we refer to Theorem 2.2 of \cite{OS4}.

 \end{document}